\documentclass[aps,prb,preprint,groupedaddress]{revtex4-1}
\usepackage{marginnote}

\usepackage{graphicx}
\usepackage{lineno}
\usepackage{subcaption}	%the new standard for subfloats
\usepackage{float}   % allows option [H]

\usepackage{amsmath,amssymb,amsfonts,amsthm,amsbsy,amstext,stmaryrd}
\usepackage{bm}% bold math

\usepackage{siunitx}		% allows the correct half-space when typing units,e.g.
\usepackage{xfrac}
\usepackage[usenames,dvipsnames,table]{xcolor}
\usepackage{url}

\newcommand{\etal}{\emph{et al.}\ }

\begin{document}

%\preprint{}

\title{Importance of elastic finite size effects: neutral defects in ionic compounds}


\author{P.A.~Burr}
\affiliation{School of Electrical Engineering and Telecommunications, University of New South Wales, Kensington, 2052, NSW, Australia.}
\email{p.burr@unsw.edu.au}
\author{M.W.D.~Cooper}
\affiliation{Materials Science and Technology Division, Los Alamos National Laboratory P.O. Box 1663, Los Alamos, NM 87545, USA.}

\date{\today}

\begin{abstract}
Small system sizes are a well known source of error in DFT calculations, yet computational constraints frequently dictate the use of small supercells, often as small as 96 atoms in oxides and compound semiconductors. %, and careful consideration of finite size effects are crucial to obtain credible predictions of defect properties.
%Small system sizes are a well known limitation of DFT calculations, particularly when considering charged defects where electrostatic interactions are paramount. On the other hand, self-interaction of neutral defects is often discounted or assumed to follow an asymptotic behaviour and thus easily countered by linear elastic theory (although rarely such corrections are applied in ionic compounds).
%Computational considerations often dictate the use of small supercell models, often as small as 96 atoms in oxides and compound semiconductors, and careful attention to finite size effects are crucial to obtain credible predictions of defect properties.
In ionic compounds, electrostatic finite size effects have been well characterised, but self-interaction of charge neutral defects is often discounted or assumed to follow an asymptotic behaviour and thus easily corrected with linear elastic theory. % (although rarely such corrections are applied in ionic compounds).
Here we show that elastic effect are also important in the description of defects in ionic compounds and can lead to qualitatively incorrect conclusions if inadequatly small supercells are used; moreover, the spurious self-interaction does not follow the behaviour predicted by linear elastic theory. %Considering the exemplar cases of metal oxides with fluorite structure, and related non-oxide compounds, we show that elastic interactions are also important in the description of defects in ionic compounds and can lead to qualitatively incorrect conclusions if insufficiently large cells are used.
Considering the exemplar cases of metal oxides with fluorite structure, we show that numerous previous studies, employing 96-atom supercells, misidentify the ground state structure of (charge neutral) Schottky defects. We show that the error is eliminated by employing larger cells (324, 768 and 1500 atoms), and careful analysis determines that elastic effects, not electrostatic, are responsible. The spurious self-interaction was also observed in non-oxide ionic compounds and irrespective of the computational method used, thereby resolving long standing discrepancies between DFT and force-field methods, % (finding agreement when compared across same supercells),
 previously attributed to the level of theory. 
%moreover these elastic effects are due to complex interactions between defect strain fields that are not described accurately by linear elastic theory.
%Additionally, our findings resolve a long standing discrepancies between DFT and force-field methods (finding agreement if compared across same supercells), previously attributed to the level of theory. %; and we show that the spurious self-interaction is observed irrespective of the method used.
%The elastic self-interaction also accounts for discrepancies between DFT and force-field methods, previously attributed to the level of theory.
The surprising magnitude of the elastic effects are a cautionary tale for defect calculations in ionic materials, particularly when employing computationally expensive methods (e.g.\ hybrid functionals) or when modelling large defect clusters.
We propose two computationally practicable methods to test the magnitude of the elastic self-interaction in any ionic system.
%performing simulations with hybrid functionals and other computationally expensive methods.
In commonly studies oxides, where electrostatic effects would be expected to be dominant, it is the elastic effects that dictate the need for larger supercells --- greater than 96 atoms.
\end{abstract}

\maketitle

%\begin{linenumbers}
\section{Introduction}

Finite size effects have been a known limitation since the beginning of atomic scale simulations of solids \cite{Catlow1982,Freysoldt2014}. 
These arise when atomic interactions extend beyond the simulation boundaries. Most modern atomic scale simulations methods adopt periodic boundary conditions (PBC) to represent crystalline matter and introduce defects through the use of supercells~\cite{Leslie1985}.
Of primary concern for finite-size effects are long range interactions through elastic (strain) fields and electrostatic (Coulomb) fields.
Elastic self-interactions of point defects have been studied since the early days of atomic scale simulations, first for simple elemental metals~\cite{Eshelby1955,Hardy1967,Gairola1978,Hardy1968}, and later for ionic crystals~\cite{Gillan1984,Leslie1985}. This was dictated by necessity, as the exhisting computational resources limited the size of force-field simulations to tens of atoms --- now $10^{12}$ atoms can be modelled \cite{Eckhardt2013}.
%Today, the same system sizes are routinely simulated using more sophisticated descriptors of the atoms, in particular DFT has become a widely used tool in for materials simulations.
%However, much of the old knowledge regarding the importance of accounting for elastic self-interactions seems to be ignored or forgotten in many recent reports.
While the computational power available to atomic scale modellers has increased dramatically, 
%enabling force-field simulations containing $10^12$ atoms \cite{Eckhardt2013},
the typical simulation size used for density functional theory (DFT) calculations has not increased accordingly, %; instead, the same simulations sizes are often used (typically between 30 and 120 atoms),
in favour of ever increasing sophistication in the description of the electronic state. %(exchange-correlation funcitonals ascending Jacob's ladder of approximation, introduction of dynamical method over static calculations)
However, much of the old knowledge regarding the importance of accounting for elastic self-interactions seems to be lost or ignored in many recent reports.

%In crystalline solids, defects interact over long ranges primarily through elastic (strain) fields and electrostatic (Coulomb) fields.
%Both elastic and electrostatic interactions are known to cause finite size effects if simulations sizes are excessively small, however
%Finite size effects are a known limitation of atomic-scale simulations and the first attempt at countering these spurious effects date to the Mott-Littleton approach of 1938~\cite{Mott1938}. Most atomic scale simulations methods adopt periodic boundary conditions (PBC) to represent crystalline matter and finite size effects arise when long-range interactions extend beyond the simulations boundaries.
%Defects in crystals primarily interact over long ranges through elastic (strain) fields and electrostatic (Coulomb) fields.
%Finite size effects may arise when the interaction field of the defect extends beyond the size of the simulation cell. {\color{red} Finite size effects are a known limitation of atomic-scale simulations and the first attempt at countering these spurious effects date to the Mott-Littleton approach of 1938~\cite{Mott1938}. Since then many alternative approaches have been devised [ref]} to counter both elastic and electrostatic finite size effects, especially for use in conjunction with periodic boundary conditions (PBC) used to simulate condensed matter.
In ionic materials, where typically point defects are charged, Coulomb interactions are assumed to be predominant.  Consequently, a large body of research has focussed on predicting and countering the electrostatic self-interaction energy in DFT, producing a number of charge correction schemes with increasing degrees of complexity and sophistication~\cite{Makov1995,Lany2009,Freysoldt2009,Hine2009,Freysoldt2011,Taylor2011,Murphy2013,Kumagai2014}. On the other hand, elastic self-interactions, which are thoroughly accounted for in metals and iono-covalent materials~\cite{Freysoldt2014,Leslie1985,Varvenne2013,Jain2016}, have largely been neglected in strongly ionic compounds, being perceived of secondary importance to electrostatic interactions. Here we show that for charge neutral defects, the elastic interactions are non-negligible and lead to a qualitative change in defect stability.

We consider exemplar cases of metal oxides with fluorite structure (CeO$_2$, ThO$_2$, UO$_2$ and actinide oxides) that have been extensively studied in the past due to their engineering applications including solid oxide fuel cells (SOFC), electrolyzer cells, ion conductors, catalysts and nuclear fuel. %We also compare the behaviour of the ionic compounds with that of an isostructural metallic compound, showing that the phenomenon is 
%Metal oxides with formula MO$_2$ and fluorite crystal structure (represented in figure~\ref{fig:SDfluorite}) have been extensively studied due their many applications as functional oxides. CeO$_2$ and Yttria-stabilised zirconia are widely used as an electrolyte in solid oxide fuel cells (SOFC), solid oxide electrolyzer cells, oxygen ion conductors and catalytic converters. UO$_2$ is used as nuclear fuel in most nuclear reactors worldwide and the iso-structural ThO$_2$, which is also a common catalytic material, is considered to be a viable alternative. PuO$_2$ and minor actinide oxides (NpO$_2$, AmO$_2$ and CmO$_2$) also exhibit fluorite structure and are present in both spent nuclear fuel and mixed oxide fuel concepts.
%\begin{figure}[hbt]
%\centering
%\marginnote{this figure might have to go to save space. I might put it on Wikipedia.}
%\includegraphics[width=0.5\textwidth]{Xtals_combined_2.pdf}
%\caption{\label{fig:fluorite} Two representations of the fluorite crystal structure: (left) the conventional unit cell and (right) a unit cell with origin shifted by $(\frac{1}{2},0,0)$. The latter will be used later in this letter.}
%\end{figure}
Specifically we consider the formation of charge-neutral Schottky clusters ($\left\{V^{\boldsymbol{\cdot\cdot}}_\mathrm{O} : V^{\prime\prime\prime\prime}_\mathrm{M} : V^{\boldsymbol{\cdot\cdot}}_\mathrm{O} \right\}^\times$), which are known to reduce oxygen mobility \cite{Genreith-Schriever2015,Zacate2000,Grieshammer2014,Grieshammer2016}, degrade the electrolytic properties of SOFC \cite{Grieshammer2016,Li2011a,Kim2012a} and govern the distribution and retention of gaseous fission products in nuclear fuel \cite{Chroneos2016,Bertolus2015,Ichinomiya2009,Cooper2014f,Liu2012}. 
%In the fluorite structure, bound Schottky defects may exhibit three possible configurations, named SD$_{[100]}$, SD$_{[110]}$ and SD$_{[111]}$, depending on the arrangement of anion vacancies around the cation vacancy (see figure~\ref{fig:SDfluorite}).
%\begin{figure}[hbt]
%\centering
%\includegraphics[width=0.495\textwidth]{SD_combined_5.png}
%\caption{\label{fig:SDfluorite} Three possible configurations of the bound SD in the fluorite structure.}
%\end{figure}
We also extend the study to non-oxide ionic compounds with fluorite structure (CaF$_2$) and anti-fluorite structure (Be$_2$C, also ionic~\cite{Herzig1985,Tzeng1998a}) to show that the phenomenon is not limited to oxides.

\begin{figure}[hbt]
\centering
\includegraphics[width=0.495\textwidth]{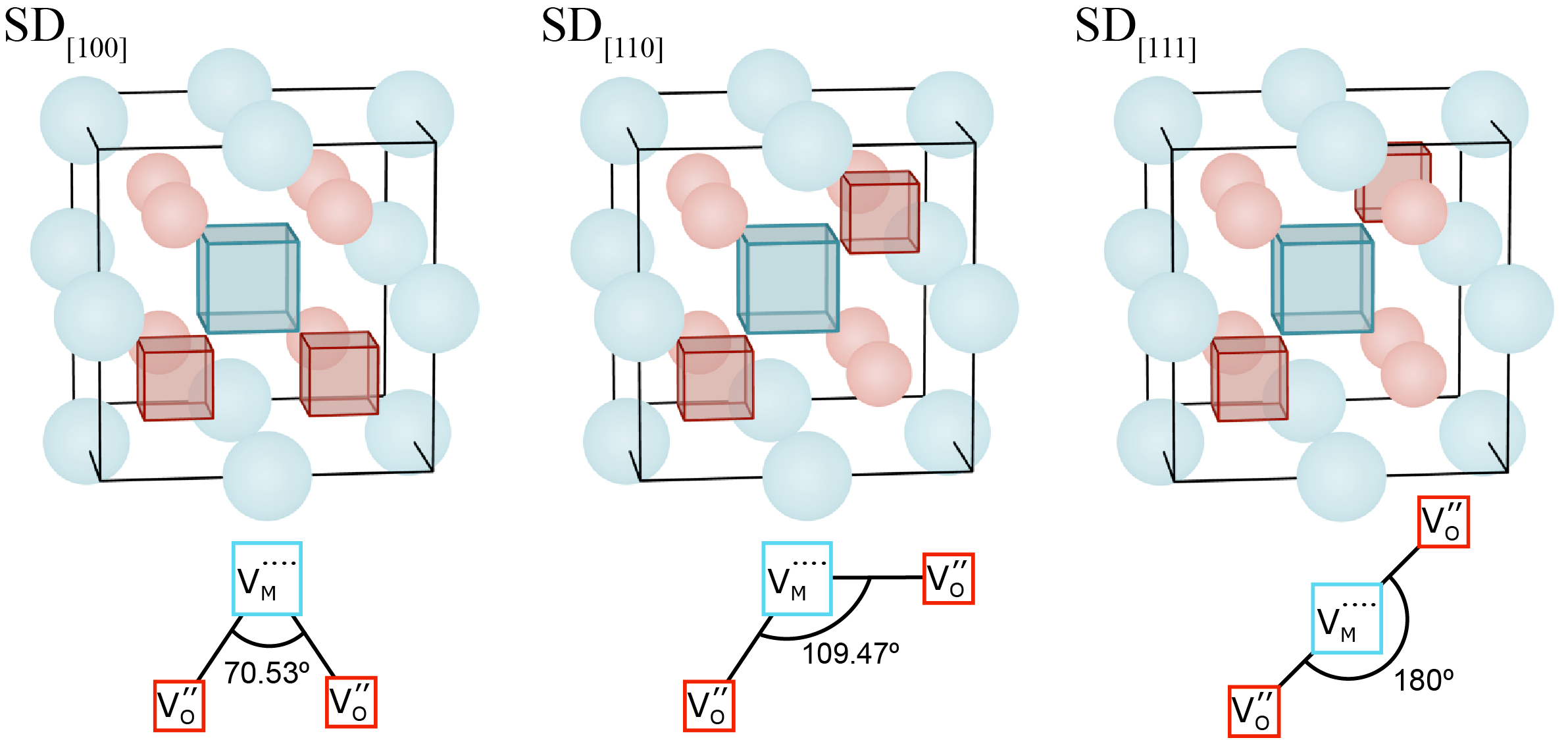}
\caption{\label{fig:SDfluorite} Three possible configurations of the bound SD in the fluorite structure, defined by the arrangement of the anion vacancies around the cation vacancy.}
\end{figure}

The formation and migration of charge-neutral clusters in these metal oxides has been extensively investigated with \emph{ab-initio} simulations \cite{Grieshammer2016,Li2011a,Liu2012,Shao2017,Murgida2014,Wang2014,Zacherle2013,Keating2013,Liu2011a,Janek2009,Xiao2011a,Gupta2009,Keating2012,Thompson2011,Crocombette2012,Crocombette2011,Dorado2013,Geng2008,Vathonne2014,Huang2014}, but nearly all of the studies were carried out with simulation cells containing up to 96 atoms ($2\times2\times2$ supercell) with few instances where larger supercells were used \cite{Vathonne2014,Bradley2015,Kuganathan2017,Murphy2014c}. 
Of the three possible configuration that a Schottky cluster may exhibit (see figure~\ref{fig:SDfluorite}), the DFT studies consistently report the the SD$_{[110]}$ configuration as the lowest energy cluster.
%In the studies concerned, the SD$_{[110]}$ configuration is consistently reported as the lowest energy cluster.
One exception is the publication by Yu \etal~\cite{Yu2009a}, however their reported Schottky energies appear to be orders of magnitude smaller than all other published work. On the other hand, modelling studies employing empirical force-fields, and thus frequently using very large periodic simulations cells or the Mott-Littleton approach~\cite{Mott1938}, often report the inverse: that SD$_{[111]}$ is more favourable than SD$_{[110]}$ \cite{Govers2007,Aidhy2009,Cooper2014} --- although, there is large variation in results as the quality of the potentials precludes the apparent stability of Schottky clusters.
%Whilst this discrepancy may be attributed in part of the different level of theory of the two methods, it is reasonable to believe that finite size effects of the periodic boundary conditions may play a significant role.
It is unclear to what extent the discrepancy between empirical and \emph{ab-initio} results is due to the different level of theory (DFT v.s.~force-fields) or to finite size effects. In this paper we show that DFT and reliable force-field potentials are indeed in agreement if the results are compared across the same supercell sizes. 
%Notably, one DFT publication by Murphy \etal \cite{Murphy2014c} on ThO$_2$  showed that by increasing the size of the simulation cell, the energy difference between SD$_{[110]}$ and SD$_{[111]}$ diminishes considerably. 
Notably, two studies employed $3\times3\times3$ supercell (324 atoms) of ThO$_2$ \cite{Kuganathan2017,Murphy2014c}. Murphy \etal \cite{Murphy2014c} report similar trends between DFT and force-field simulations for increasing supercell size, however the discrepancy regarding the most favourable cluster configuration still remained. Although not belonging to the family of fluorite compounds, Bradley \etal~\cite{Bradley2015} also employed a 324-atoms supercells of \emph{m}-HfO$_2$ to study defect clusters.
%Notably, three studies employed $3\times3\times3$ supercells (324 atoms) \cite{Bradley2015,Kuganathan2017,Murphy2014c}. Murphy \etal \cite{Murphy2014c} report similar trends between DFT and force-field simulations for increasing supercell size, however the discrepancy regarding the most favourable cluster configuration still remained.

In this letter, we show that charge neutral Schottky clusters interact over long ranges through elastic fields, and that the commonly used 96-atom supercell is inadequately small to capture the correct ground state of Schottky clusters.
We then show that the root cause of this finite size effect is spurious interaction between strain fields across PBC, and not electrostatic or electronic effects. Additionally, the elastic finite size effect is not described accurately by linear elastic theory, owing to the complex mode of relaxation. We conclude by proposing two methods to estimate, albeit approximately, the magnitude of the elastic self-interaction for any ionic system.
%The extent of this finite size effect is such that all previous ab-initio computational efforts on fluorite oxides failed to captured the correct cluster configuration as the most energetically favourable.
%We also show for the first time, agreement between force-field simulations and DFT simulations, without fitting the empirical potentials to DFT configurations. Furthermore, we show that our results are not sensitive to the choice of computational methodology (and the various flavours within DFT), and that the root cause of this finite size effect is spurious interaction between strain fields across PBC. %destructive interference of strain fields across PBC.
%for this apparent inversion of most favourable cluster is the constrains on atomic relaxation caused by the PBC.
%Lastly we show that this peculiar finite size effect is not limited to MO$_2$ compounds with fluorite structure, but it is also observed in non-oxide compounds with fluorite structure (CaF$_2$) or anti-fluorite structure (Be$_2$C).

%\end{linenumbers}
\section{Methodology}

Force-field simulations were performed with the LAMMPS code \cite{LAMMPS,Plimpton1995}, using the many-body potential of Cooper, Rushton and Grimes (CRG) \cite{Cooper2014}, as this potential set proved to be reliable and transferable \cite{Ghosh2016,Liu2016,Cooper2016a,Yuan2015a,Rushton2014} across a wide range of MO$_2$ compounds.
DFT simulations were carried out with VASP \cite{Kresse1996a}, using the PBE exchange-correlation functional \cite{Perdew1996}, PAW pseudo-potentials with the maximum number of valence electrons available \cite{Kresse1999} and a plane-wave cut-off of \SI{500}{eV}. Details of the k-point grids and evidence of convergence to within \SI{1}{meV} is provided in the supplementary material\cite{SupplMater}.
$c$-ZrO$_2$ was also initially investigated, but due to the stability of $m$-ZrO$_2$ it was not possible to retain the cubic symmetry during relaxation of defects in large cells, even by constraining some degrees of freedom.

On-site Coulomb correction terms have been included for CeO$_2$ and UO$_2$, following the majority of the published literature \cite{Allen2014,Keating2012,Andersson2007,Wiktor2017,Vathonne2014,Dorado2013,Dorado2013a,Thompson2011,Crocombette2012,Crocombette2011,Dorado2010,Geng2008,Iwasawa2006,Kotani1992,Yamazaki1991}: Dudarev \etal's formalism~\cite{Dudarev1998a} for CeO$_2$ with $U\{\text{Ce}_{4f}\} = 5.0\text{ eV}$, $U\{\text{O}_{2p}\} = 5.5\text{ eV}$; and Liechtenstein's formalism \cite{Liechtenstein1995} for UO$_2$ with $U\{\text{U}_{5f}\} = 4.5\text{ eV}$  and $J\{\text{U}_{5f}\} = 0.51\text{ eV}$.
%The simulations of CeO$_2$ were repeated with no +U yielding qualitative agreement, presented in the supplementary materials.
U-ramping \cite{Meredig2010} (for CeO$_2$) and occupation matrix control \cite{Allen2014,Dorado2010} (for UO$_2$) were used to avoid metastable states. Details are provided in the supplementary material~\cite{SupplMater} together with results obtained without +U to emphasis that the findings are not sensitive to the choice of simulation parameters.
UO$_2$ was described with collinear 1{\bf k} antiferromagnetic ordering, as this is the best collinear approximation of the true (non-collinear 3{\bf k} AFM \cite{Desgranges2016,Ikushima2001}) magnetic ordering of UO$_2$ \cite{Wiktor2017,Geng2008,Vathonne2014,Dorado2013,Dorado2013a,Crocombette2012,Crocombette2011,Dorado2010}.
In ionic materials, the formation energy of a defect $d$ with charge $q$ is conventionally calculated as
\begin{equation}
%E^f_d = E(\text{defect}) - E(\text{perfect}) \pm \sum_\alpha n_\alpha \mu_\alpha - q \mu_e + E_{corr}
E^f_d = E_\text{def} - E_\text{perf} \pm \sum_\alpha n_\alpha \mu_\alpha - q \mu_e + E_\text{chcor} + E_\text{elcorr}
\end{equation}
where $E_\text{def}$ and $E_\text{perf}$ is the DFT total energies of the defective or pristine supercells, $\mu_\alpha$ is the chemical potential of all species added or removed to form the defect, $\mu_e$ is the fermi level of the system normalised by the valence band maximum, $E_\text{chcorr}$ is a charge correction term, following a number of possible schemes \cite{Makov1995,Lany2009,Freysoldt2009,Freysoldt2011,Taylor2011,Hine2009,Murphy2013,Kumagai2014}, and $E_\text{elcorr}$ is the energy due to elastic self-interaction, seldom accounted for in ionic compounds. Since (a) the defects considered here are charge neutral, (b) the composition of the defects is precisely one stoichiometric formula unit (i.e.\ $\mu_{MO_2} = \frac{E_\text{perf}}{x}$, where $x$ the number of formula units in the supercell), and (c) the elastic self-interaction is the subject of the study, the defect formation energy of a Schottky cluster is simplified to:
\begin{align}
%E^f_\text{SD}	=& E(\text{defect}) - E(\text{perfect}) - \frac{E(\text{perfect})}{x}\\
E^f_\text{SD}	=& E_\text{def} - E_\text{perf} - \frac{E_\text{perf}}{x}\\
E^f_\text{SD}	=& E(\mathrm{M}_{x-1}\mathrm{O}_{2x-2}) - \frac{x-1}{x} E(\mathrm{M}_x\mathrm{O}_{2x})
\end{align}
%\begin{equation}
%E^f_\text{SD} = E(\mathrm{M}_{x-1}\mathrm{O}_{2x-2}) - \frac{x-1}{x} E(\mathrm{M}_x\mathrm{O}_{2x}) 
%\end{equation}
%\begin{equation}
%E^f_\text{SD} = E(\text{defect}) - E(\text{perfect}) = E(\mathrm{M}_{x-1}\mathrm{O}_{2x-2}) - \frac{x-1}{x} E(\mathrm{M}_x\mathrm{O}_{2x}) 
%\end{equation}
in line with previous publications \cite{Thompson2011,Crocombette2012,Crocombette2011} (except for the explicit inclusion of elastic self-interaction term). This simplification conveniently removes any dependence of our results from external factors such as chemical potential of reference elements, or apparent bang gap of the material.

%In metals and covalent compounds, $E_\text{elcorr}$ can be estimated from the elastic dipole of the defect~\cite{Varvenne2013}., with the aid of the \emph{aneto} script and using elastic constants obtained from DFT simulations through lattice perturbation to retain self-consistency.
The linear elastic theory approximation to $E_\text{elcorr}$ was calculated with the aid of the \emph{aneto} script \cite{Varvenne2013} from the stress tensor of the relaxed simulations and using elastic constants obtained from DFT simulations through lattice perturbation to retain self-consistency.
Selected calculations were repeated where atomic relaxation was restricted to atoms within a \emph{relaxation radius} from the defect centre, and all other atoms were kept fixed at the perfect lattice site. %This was achieved through the ``selective dynamics'' option of VASP, in conjunction with a python script, provided in the supplementary materials.

\section{Results and discussion}

\begin{figure*}
\centering
\includegraphics[width=\textwidth]{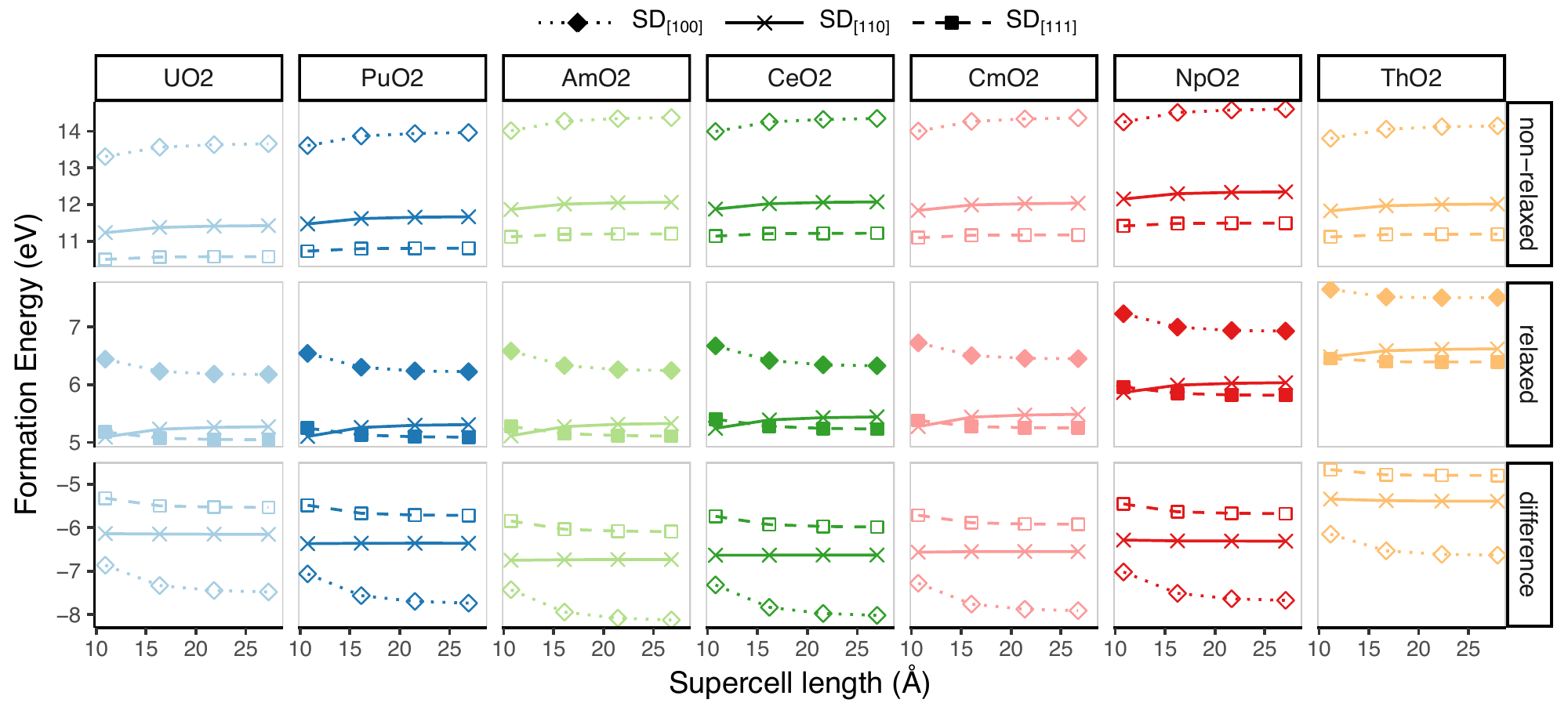}
\caption{\label{fig:supercellsEAM} Defect formation energy (before and after relaxation) from CRG potentials versus supercell size (form $2\times2\times2$ to $5\times5\times5$) for three bound Schottky configurations in actinide oxides and CeO$_2$. Values for ThO$_2$ were taken from~\cite{Murphy2014c}.}
\end{figure*}

The effect of finite PBC was first investigated using the CRG potential. Figure~\ref{fig:supercellsEAM} shows the formation energy (filled symbols) of bound Schottky clusters in various actinide oxides and CeO$_2$. It is clear that, irrespective of the cation species, the SD$_{[111]}$ defect is the most favourable when simulated in large enough supercells (containing 324 atoms or more), but the SD$_{[110]}$ is the most favourable in the smaller simulation cell containing 96 atoms.
%The only exception to this is ThO$_2$, which exhibits SD$_{[111]}$ as the most favourable configuration even in the $2\times2\times2$ supercell, but still shows the same trends in SD formation energy as a function of supercell size. 
The results of calculations before geometry relaxation is also presented in the top panel, showing no cross-over of defect energies.

\begin{figure}
\centering 
\includegraphics[width=0.495\textwidth]{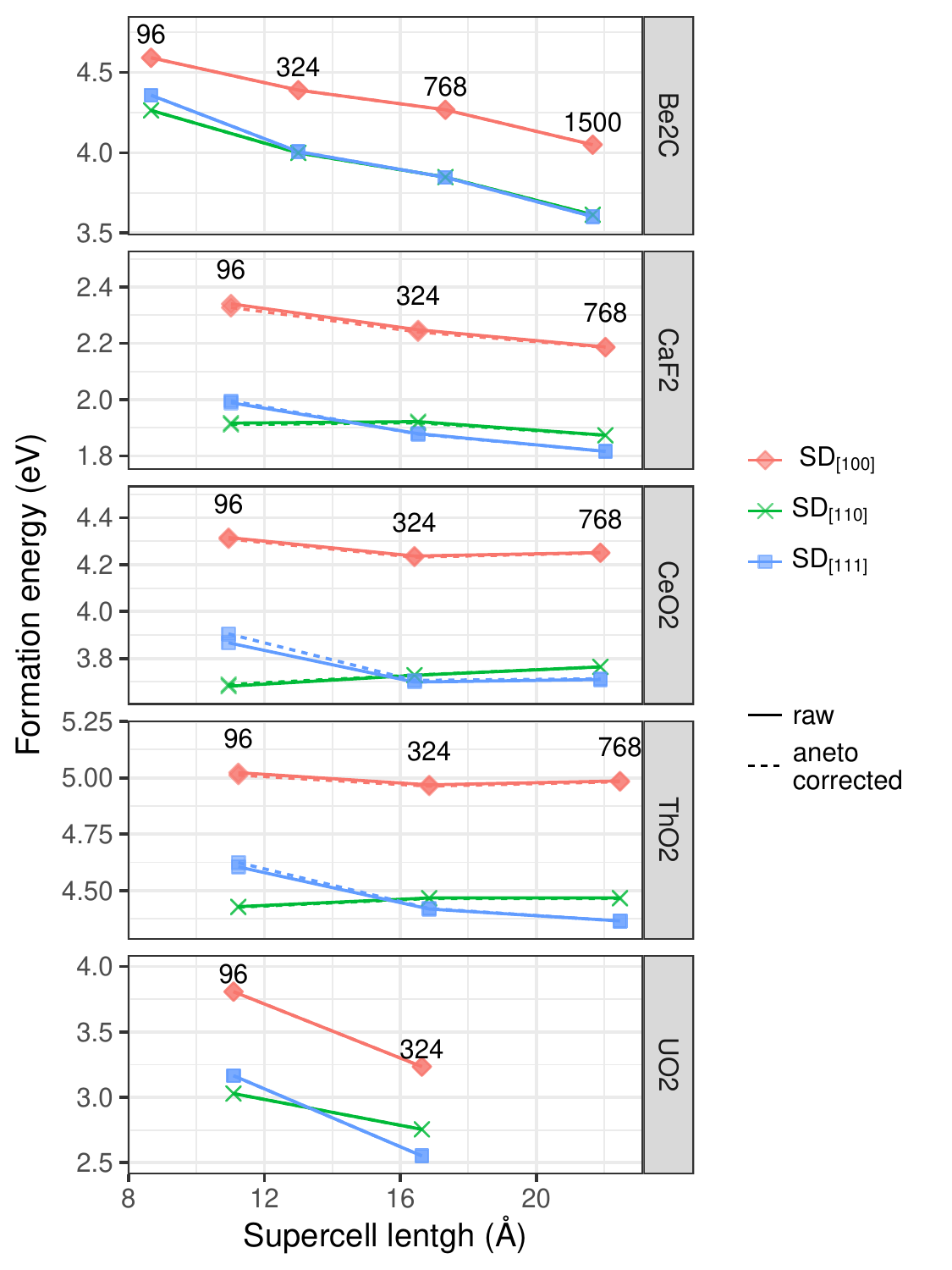}
\caption{\label{fig:supercellsDFT} Defect formation energies from DFT as a function of supercell size (labels indicate atoms in supercell). UO$_2$ simulations were limited to 324 atoms, due to the additional complexity and computational cost of OMC.}
\end{figure}

To show that the crossover was not a peculiarity of the CRG potential form, DFT calculations were performed with supercells containing 96, 324 and 768 atoms on selected oxides (CeO$_2$, ThO$_2$, UO$_2$) as well as CaF$_2$ and Be$_2$C (Figure~\ref{fig:supercellsDFT}). Be$_2$C has considerably small lattice parameter, therefore a further supercell containing 1500 atoms was also considered. It is evident that DFT and force-field calculations are in agreement, and crossover between SD$_{[111]}$ and SD$_{[110]}$ occurs between the \SI{\sim10}{\angstrom} supercells (96 atoms) and the \SI{\sim16}{\angstrom} supercells (324 atoms). The trend is predicted for CaF$_2$ and Be$_2$C as well as the oxides. Importantly, including the energy penalty predicted from linear elastic theory (dashed lines) does not correct the trend. 

\begin{figure}
\centering
\includegraphics[width=0.495\textwidth]{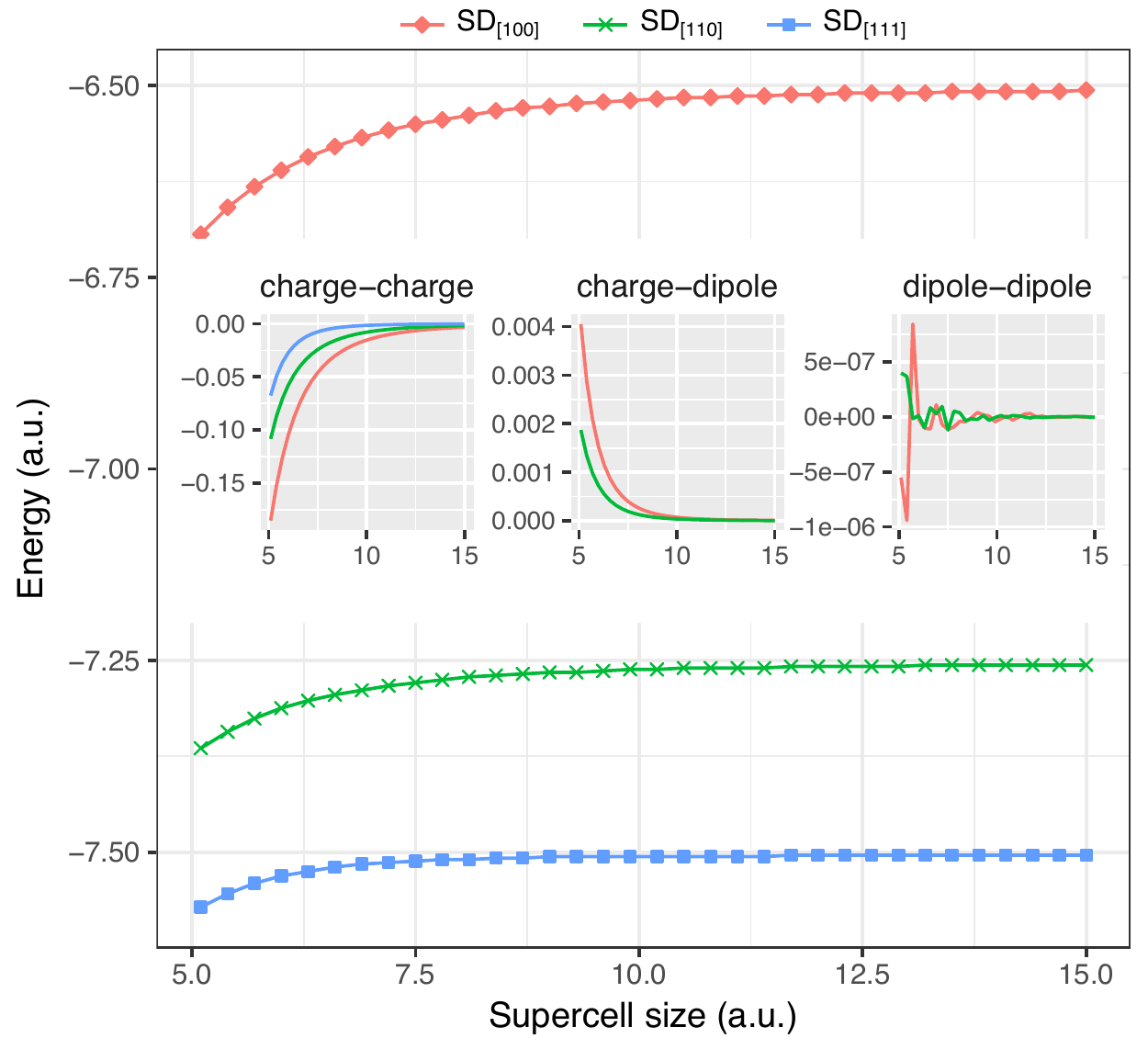}
\caption{\label{fig:electrostatics} Total electrostatic energy of SD modelled as point charges arranged in a dielectric medium with periodic boundaries. Insert: energy contribution from self-interaction across supercell boundaries.}
\end{figure}

It is well known that point defects in ionic materials interact chiefly though their charges, and although the Schottky clusters have no overall charge, they are still comprised of three charged defects ($\text{V}_{\text{anion}}^{2-}-\text{V}_{\text{cation}}^{4+}-\text{V}_{\text{anion}}^{2-}$) that may individually interact across periodic boundaries. In addition, SD$_{[100]}$ and SD$_{[110]}$ also exhibit effective dipoles since the geometrical centre of the positive charges does not align with that of the negative charges. SD$_{[111]}$ does not have an associated dipole since it is a linear defect with mirror symmetry.
%Because of this, one might suspect that dipoles play an important role in this peculiar finite size effect. However, we find that this is not the case by evaluating all charge-charge, charge-dipole, and dipole-dipole interactions independently (Figure~\ref{fig:electrostatics}).
Nevertheless, to show that electrostatic interactions alone cannot account for this peculiar finite size effect, all charge-charge, charge-dipole, and dipole-dipole interactions have been evaluated independently (Figure~\ref{fig:electrostatics}).
Three point charges (two positive and one negative with no overall charge) have been arranged in a dielectric medium with the same configurations as the three bound Schottky defects, and then replicated in a repeating array of $10 \times 10 \times 10$ to model the effect of PBC, where the distance between replicas was increased progressively to simulate larger supercells. Dielectric constants, lattice parameter and the magnitude of charges are arbitrary. However, a range of $\frac{q^2}{\varepsilon a}$ ratios have been modelled, all yielding the same qualitative behaviour.

Figure~\ref{fig:electrostatics} shows that as the supercell size increases, the energy contribution from self-interaction across periodic boundaries tend to zero (see insets), therefore the total energy of each system converges toward the internal energy of the point charge triplet, consisting entirely of Coulomb interactions. 
It is clear that the three point charge configurations (representative of the three Schottky clusters) never cross over at any separation, hence, the electrostatic interactions alone do no account for the change in relative stability of Schottky clusters. 
This is reassuring, given that CaF$_2$ shows similar trend to CeO$_2$, ThO$_2$ and UO$_2$ despite a factor of $\frac{1}{2}$ difference in ionic charges (Figure~\ref{fig:supercellsDFT}).
%The findings are also consistent with the behaviour of defects in un-relaxed fluorite structure (Figures~\ref{fig:supercellsEAM}~and~\ref{fig:supercellsDFT}). Notably the anti-fluorite compound Be$_2$C exhibits opposite behaviour in the unrelaxed system.

Beyond electrostatic effects, electronic effects may also lead to a change in behaviour with increasing supercell size. This has been well documented for charge neutral vacancies in Si~\cite{Mercer1998,Puska1998,Centoni2005}, where small simulation cells (64-atoms or fewer) predict a retention of $T_d$ symmetry, while larger cells correctly identify a reduction of local symmetry to $D_{2d}$, as observed experimentally. The symmetry reduction, and associated energy reduction, is due to a Jahn-Teller distortion~\cite{Jahn1957}, and consequently the effect is strongly sensitive to Brillouin zone sampling as well as supercell size \cite{Puska1998}. SiC is another example of covalent material where similar effects have been observed~\cite{Zywietz1999}. Jahn-Teller effects are not limited to covalently bonded materials, in fact they are known to be important in many oxides~\cite{Dorado2010,Elfimov1999}. However, the finite-size effect observed in the current study was consistently reproduced with GGA, GGA+U and force field potentials (with fixed charges on atoms), suggesting that electronic effects cannot be at the heart of the matter.

%Change in behaviour with supercell size may be caused also by electronic effects, as documented in the well studied case of charge-neutral vacancies in Si.

\begin{figure*}[hbt]
\centering
\includegraphics[width=\textwidth]{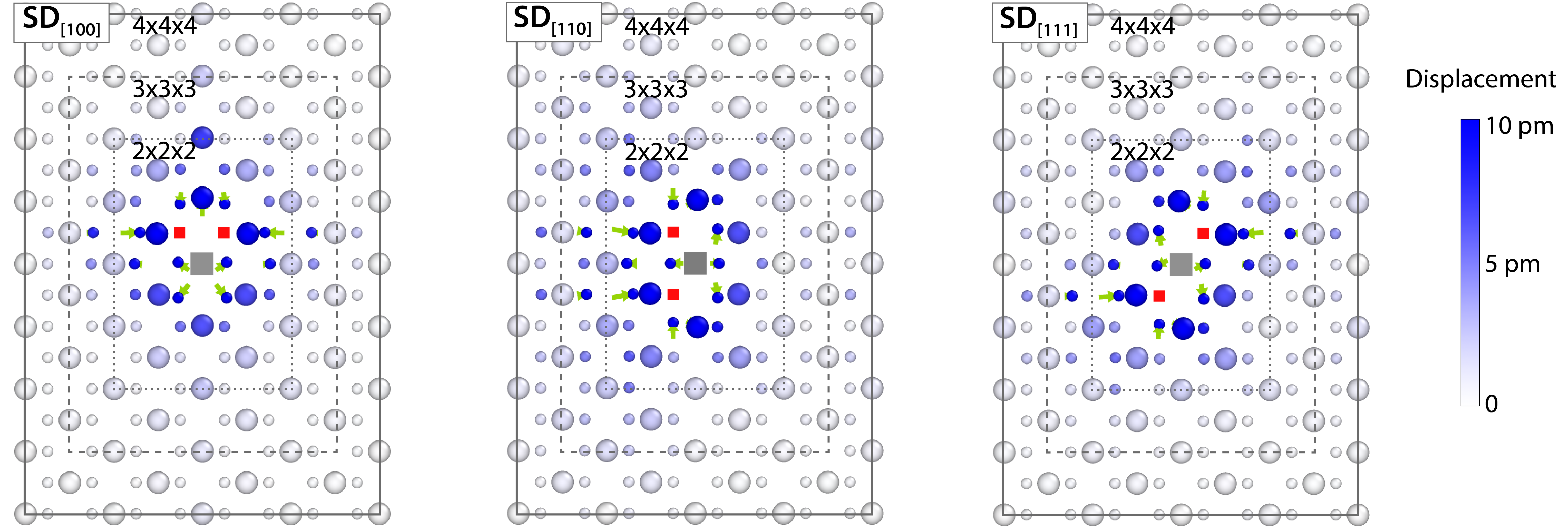}
\caption{\label{fig:strainfields} (110) slice (three atomic layers) of the 768-atoms supercell of CeO$_2$.
%(110) slice of the atomic displacements in the 768-atoms supercell of CeO$_2$ arising from bound Schottky clusters. For clarity, only three atomic layers are shown.
Dashed lines represents the boundaries of the smaller supercells.
Grey and red squares represent the Ce and O vacancies respectively, green arrow represent the displacement magnitude scaled up by a factor of 5.
First nearest neighbour atoms exceed the colour map with displacements of up to \SI{26}{pm}.}
\end{figure*}

\begin{figure}[hbt]
\centering
\includegraphics[width=0.495\textwidth]{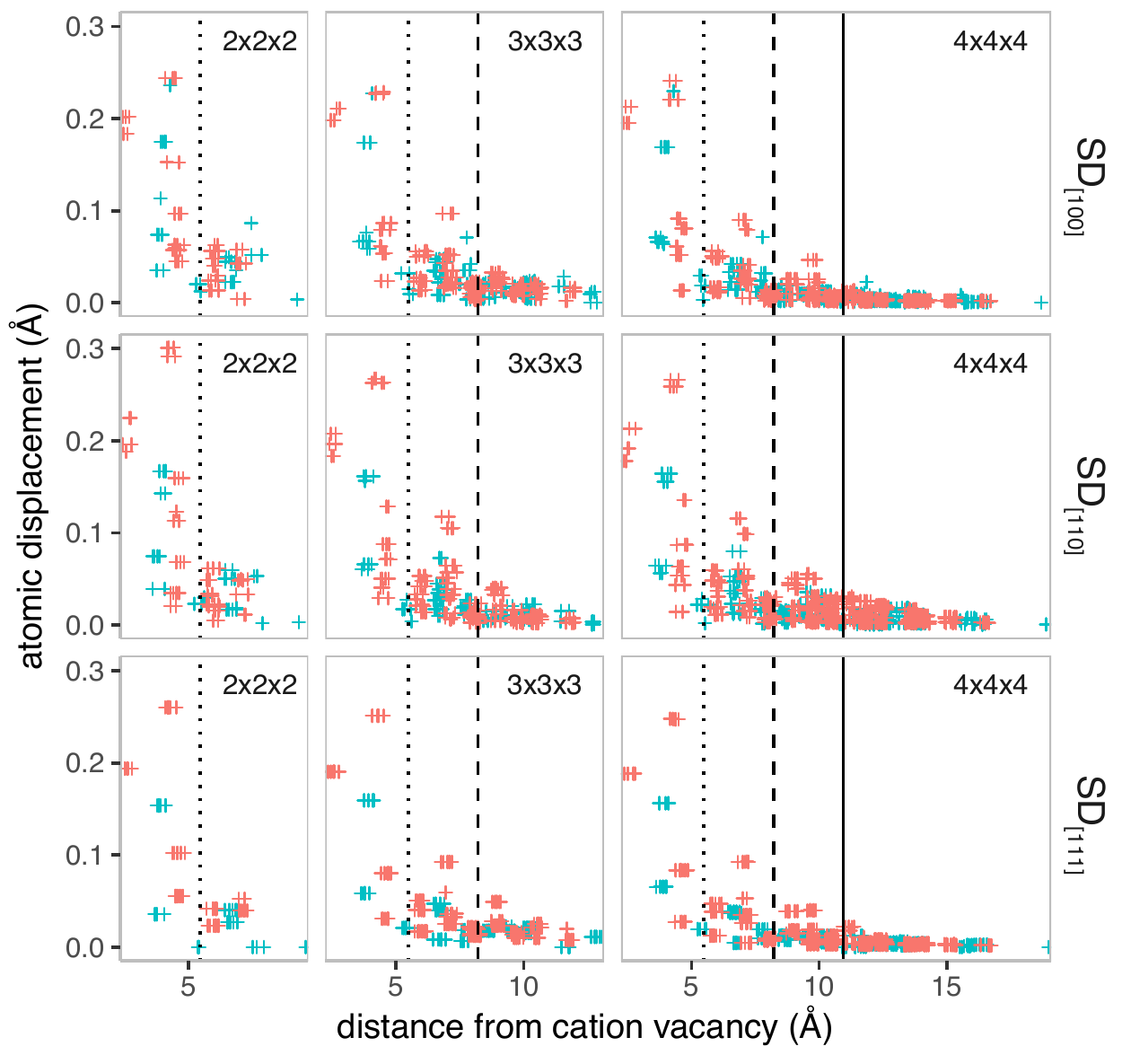}
\caption{\label{fig:displacements} Atomic displacements caused by Schottky clusters in CeO$_2$ supercells. Red and turquoise crosses represent O and Ce atoms respectively. Black vertical lines represent half of the supercell length, following the same key of Figure~\ref{fig:strainfields}. Beyond those lines, some atoms are closer to the periodic replica of the defect than the central defect.}
\end{figure}

The source of the cross-over is instead found in the elastic interactions. 
Figures~\ref{fig:supercellsEAM} shows that prior to geometry relaxation there is no cross-over in stability between SD$_{[110]}$ and SD$_{[111]}$. This is also shown for DFT calculations in the supplementary material~\cite{SupplMater2}.
Thus relaxation is hampered in the smallest supercell.
%It is therefore evident that the root cause of the spurious effect lies in artificial constrains on the ionic relaxation.
% the results of the DFT calculations prior to geometry relaxation, i.e.\ after the first self-consistent field loop, exhibiting no cross-over  in stability between SD$_{[110]}$ and SD$_{[111]}$.
%In the smallest supercell the elastic strain fields of the defects must interact significantly with their periodic replica, preventing some atoms from re-arranging into a more favourable configuration.
%It is also evident that geometry relaxation affects the SD$_{[111]}$ defect more than the other two SD configurations, which are more compact.
%By way of example, the atomic displacements cause by the three bound Schottky clusters in CeO$_2$ are shown in Figure~\ref{fig:strainfields} and are plotted in as a function of distance from the cation vacancy in Figure~\ref{fig:displacements}. 
Figure~\ref{fig:strainfields} depicts the atomic displacements caused by the three bound Schottky clusters in the largest DFT cell of CeO$_2$.
%For clarity, the atomic structures of Figure~\ref{fig:strainfields} are restricted to the $(110)$ plane containing the defect and one atomic plane above and below it. 
The boundaries of the smaller supercells are superimposed on the image to highlight that the strain field exceeds the bounds of the $2 \times 2 \times 2$ supercell (96 atoms).
The displacement fields are better quantified in Figure~\ref{fig:displacements}, where the atomic displacements are plotted as a function of distance from the cation vacancy. The atomic displacements obtained from the largest supercell, reveal that at a distance of \SI{7}{\angstrom} from $V_\text{Ce}^{\prime\prime\prime\prime}$ (i.e.\ between the $2\times2\times2$ and  $3\times3\times3$ boundaries) O atoms are being displaced by as much as \SI{11.5}{pm}, which is not insignificant. More importantly, when comparing the atomic displacements within the first \SI{10}{\angstrom} across different supercell sizes, it is evident that the fingerprint of atomic relaxation in the 96-atom supercell is fundamentally different from that of the larger supercells, i.e.\ not only the magnitude but also the \emph{shape} of the strain field is different. On the other hand, the fingerprint of atomic displacements changes only marginally when increasing supercell size further.

The fact that the shape of the displacement fields are fundamentally different between the 96-atom supercell and the larger supercells indicates that atomic relaxation is hampered by artificial restoring forces stemming from the PBC. This frustration of atomic relaxation can only provide a positive contribution to the total energy of the system.

Generally, it is possible to predict the energy contribution arising from elastic self-interaction through methods based on linear elastic theory, which combining the elastic dipole tensor of the simulation with the elastic constants of the material \cite{Leslie1985,Varvenne2013}. However, Figure~\ref{fig:supercellsDFT} shows that the \emph{aneto} correction, which has been proven successful in a variety of metallic and covalent systems \cite{Varvenne2013,Burr2015,Agarwal2016,Garnier2014a,Murali2015,Varvenne2014}, cannot counter the finite size effects observed here. %In the supplementary material we show that for a rare case of metallic compound with fluorite structure (Al$_2$Au), the correction from linear elastic theory is noticeable, unlike the ionic systems considered here.

\begin{figure}[hbt]
\centering
%\marginnote{figure needs editing}
\includegraphics[width=0.495\textwidth]{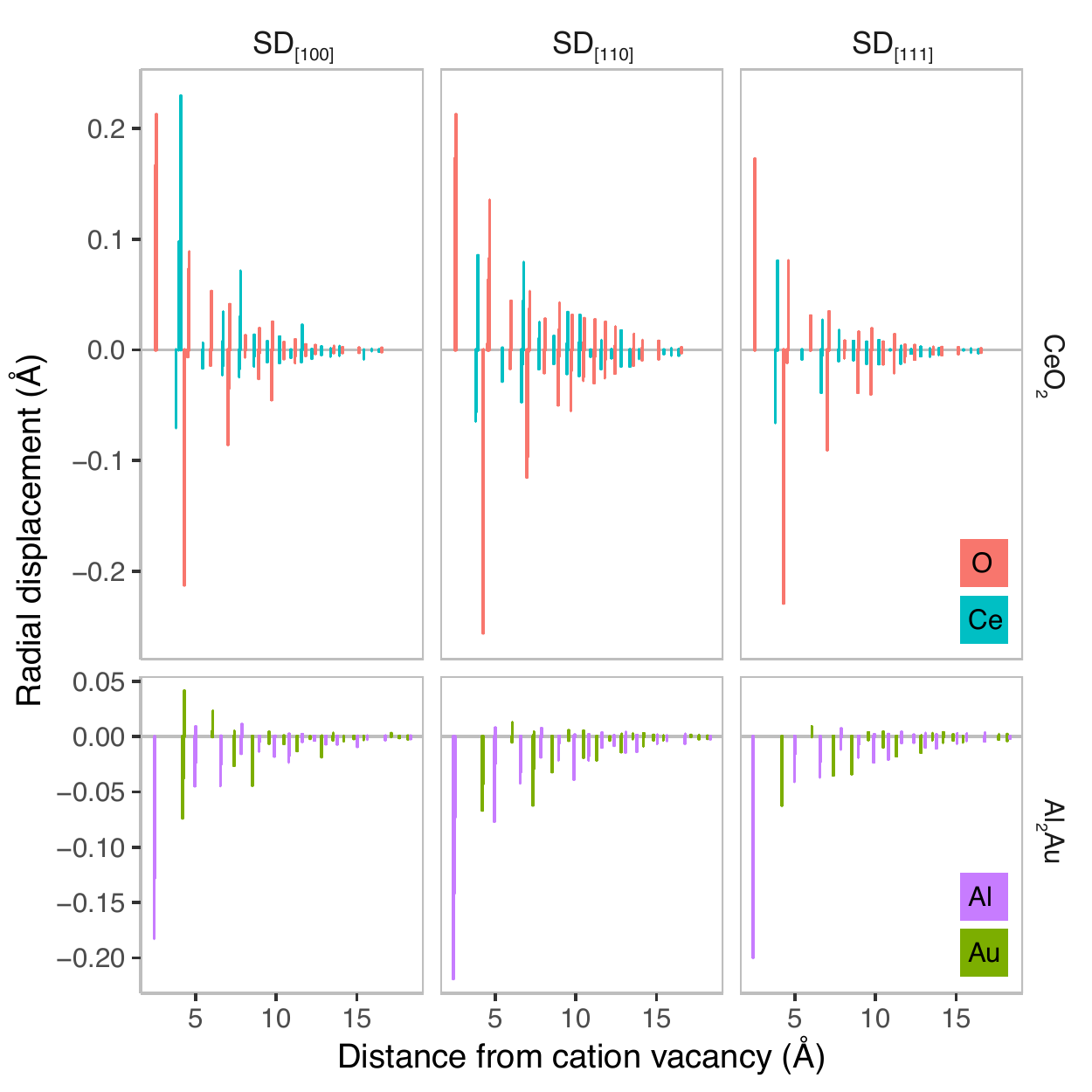}
\caption{\label{fig:radialdispl} %Radial displacement, $d_r$, of atoms surrounding an SD$_{[110]}$ cluster in CeO$_2$ and Al$_2$Au.
Radial displacement, $d_r$, caused by Schottky clusters in CeO$_2$ and Al$_2$Au. $d_r = \bm{d} \cdot \bm{r}$, where $\bm{d}$ is the displacement vector and $\bm{r}$ is the position vector with respect to the cation vacancy. %Similar plot were produced for the other two Schottky cluster configurations, yielding equivalent results (see supplementary material).
}
\end{figure}

The inability of linear elastic theory to predict the energy contribution in ionic compounds is attributed to the complex relaxation pattern caused by the defect. Figure~\ref{fig:strainfields} clearly shows that the pattern of atomic displacements is not consistent with a simple compression (acoustic) wave, where all atoms move coherently towards the vacancies to accomodate the defect volume.
%not a coherent movement of all ions towards the vacancies to accomodate a simple volume compression.
The relaxation field more closely resembles of an optical mode, where anions and cations exhibit distinct and more complex displacement patterns. 
This is evidenced more clearly in Figure~\ref{fig:radialdispl}, where the radial component of the displacement vector is compared against that of an isostructural metallic compound (Al$_2$Au). In the metallic compound, the atomic relaxations are nearly exclusively towards the vacancy cluster (negative radial displacements), and the effect diminishes nearly monotonically with distance. In the ionic compound, each shell of anions and cations exhibit both inwards and outward displacements (with the exception of the first oxygen shell).
%{\color{red} As a result, atoms that would have been equidistant from the defect centre, have spread to create narrow bands: atoms with negative displacements are correspondingly closer to the centre, thereby shifting left in the $x$-axis of Figure~\ref{fig:radialdispl}.}
As with any optical mode, this behaviour may only arise in compound materials, explaining why it has not been observed in well studied elemental materials.
In addition, Figure~\ref{fig:radialdispl} shows that the phenomenon is restricted to ionic compounds, where alternating shells of anions and cations move in opposite directions in response to displacement of charge (e.g.~neighbouring ions). %For instance, when the first shell of oxygen ions is displaced outwards in response to the repulsion caused by the cation vacancy, it affects the equilibrium forces on the next neighbouring shell of anions by increasing the repulsive force only, but a mixture of repulsive and attractive forces on the neighbouring cation shell.

%The rational for the complex and long-range relaxation pattern lies in the local electrostatic interactions between anions and cations: an outward motion of a shell of anions causes
%as the first shell of oxygen ions is repelled by the cation vacancy (and attracted by the oxygen vacancies), it moves outwards, thus causing a local perturbation in the second shell of ions (attractive for cations and repulsive for anions).
%An outwards displacement of the first shell of anions, will cause a locally attractive force towards the  oxygen ions (formally $2-$ charge) are repelled by cation vacancy ($4-$ charge), and attracted by the oxygen vacancies ($2+$ charge), and \emph{vice-versa} for the cations. The dispalcement of the first shell of ions 

%In small supercell, this complex relaxation fields may interact in ways that linear elastic theory cannot predict.
The spurious interaction results in a change in shape of relaxation near the defect core, which cannot be captured by the dipole tensor: the dipole tensor may be obtained from (\emph{a}) the (anisotropic) stress tensor and/or strain tensors on the cell, or (\emph{b}) by convergent summation of atomic displacements and/or restoring forces on atoms \cite{Nazarov2016}. In both cases the dipole tensor may only capture a truncation of the strain field, and not a change in shape of the core of the field (i.e.\ the ``optical relaxation''), which is illustrated particularly clearly in Figure~\ref{fig:displacements} by comparison between the $2\times2\times2$ and $3\times3\times3$ supercells for which there is a notable change in relaxation displacements for atoms even within \SI{5}{\angstrom} of the defect. These core atoms are \SI{<6}{\angstrom} away from the defect replica in the $2\times2\times2$ supercell.

\begin{figure}[hbt]
\centering
\includegraphics[width=0.495\textwidth]{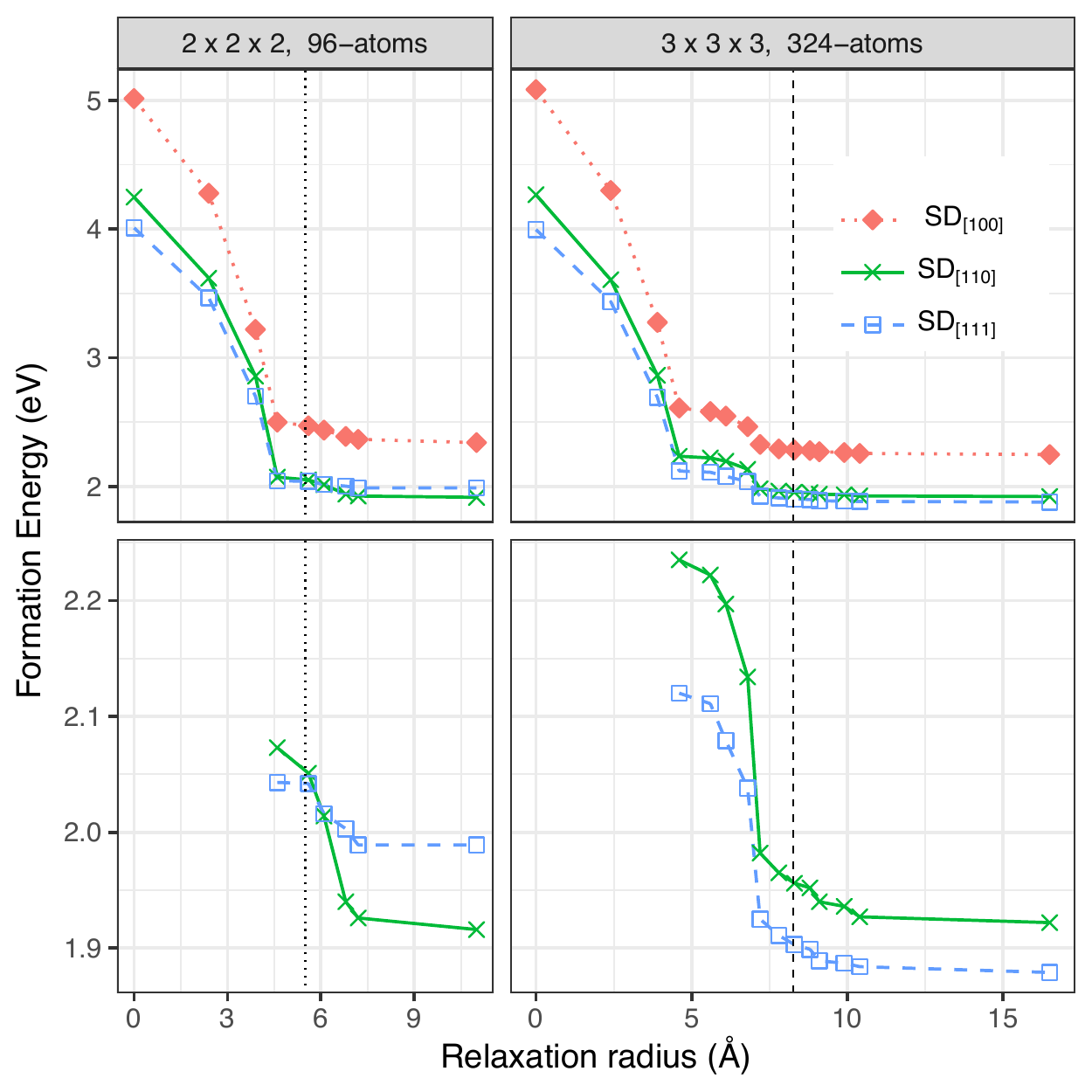}
\caption{\label{fig:seldyn} Formation energies of Schottky defects in CaF$_2$ as a function of relaxation radius, whereby all atoms beyond that radius are forcefully kept fixed in their perfect lattice sites. Lower panels provide an enlarged view of the crossover region. Vertical black line represent the largest radius that may be accommodated in the supercell without overlap $(R_\text{relax}=\frac{L}{2})$. First and last point in each plot represent a single point calculation and a complete relaxation, respectively.}
\end{figure}

As further evidence that the source of the self-interaction energy is the inhibition of elastic relaxation, we have performed calculations where only atoms within a given radius of the defect centre were allowed to relax while all other atoms were ``clamped'' at perfect lattice positions, see Figure~\ref{fig:seldyn}.
Provided that the relaxation radius is less then half the supercell length, $R_\text{relax}=\frac{L}{2}$, this constraint ensures that no strain is transmitted across the PBC.
%This two-region technique ensures that when the relaxation radius is less then half the supercell length, $R_\text{relax}=\frac{L}{2}$, no strain is transmitted across the PBC.
Thus the relaxation field, although artificially truncated at the given radius, is entirely due to the defect and not its periodic replicas.

%\begin{figure}[hbt]
%\centering
%\includegraphics[width=0.33\textwidth]{seldyn_F333_rwb_alph.png}
%\caption{\label{fig:seldyn_radius} Concentric atomic layers colour coded by radial distance from cation vacancy (grey square). White atoms (middle of colour scale) represent a radial distance of half supercell side length (\SI{8.24}{\angstrom} in this example).}
%\end{figure}

%When the relaxation radius is less then half the supercell length, $r=\fract{L}/2}$
When the relaxation radius is less then half the supercell length (i.e.\ the region of atoms that are allowed to relax fits entirely within the supercell), SD$_{[110]}$ appears less favourable than SD$_{[111]}$, consistent with the findings obtained from very large supercell simulations. Only when the relaxation radius is larger than half the cell length (i.e.\ atoms within the sphere are responding to the strain field of the defect \emph{and} its replicas), does the apparent cross-over in stability manifest in the 96-atom supercell. This is a further confirmation that the phenomenon is due to the interaction of strain fields across PBC in the smaller supercells.

Constraining atomic positions as a function of distance from the defect allows one to isolate the relaxation strain component of the defect energy. And by iterating that procedure over increasing relaxation radii, it is evident that the behaviour is distinctly not monotonic, with two clear steps at the $4^{th}$ and and $8^{th}$ nearest neighbour. This is at odds with the behaviour expected from a continuum elastic medium.%, reinforcing that linear elastic theory cannot accurately describe this elastic self-interaciton.

%{\color{red} It is evident that this two-region method (with $R_\text{relax}=\frac{L}{2}$) enables retention of a qualitative description of the defect in question, even when the supercell dimensions are inadequately small.
%The concept of a two-region method is not new to atomic scale simulations~\cite{Mott1938,Catlow1982,Lidiard1989,Grimes1990e}. However, the proposed scheme does not replace periodic boundaries, instead it is used in conjunction with PBC thus retaining their benefits with regards to the description of crystals and the use of plane waves.
%%The scheme may be easily implemented in existing PBC-based DFT codes, and it retains the benefits of PBC with regards of the use of plane waves to describe the crystal.
%Incidentally, this method also significantly reduces the computational cost of energy minimisation on account of the reduced internal degrees of freedom. Consequently, the scheme may enable the use of a larger supercell which would have otherwise been impracticable, thus potentially increasing the quantitative accuracy of the work.
%%As a side point, restricting the degrees of freedom of the atoms furthest from the defect centre increases computational speed
%}

It is concluded that if inadequately small supercells are used, 96 atoms or less for fluorite structure oxides, the spurious self interaction is beyond the reach of linear elastic theory, and thus cannot be corrected for without further calculations.
It is not always possible to increase the simulation size, especially when performing calculations with computationally expensive methods, such as hybrid functionals, ab-initio molecular dynamics or time-dependent DFT. Thus, here we propose two computationally practicable methods to ascertain whether finite size effects are significant in the system of interest:
\begin{enumerate}
\item One option is to use a lower level of theory method to perform the convergence test.
In the current work, we have shown that CRG force-field potential was as predictive as DFT in the analysis of the spurious self-interaction. Thus, if reliable force-field potentials are for the system of interest, these may be used to test supercell convergence. Similarly, for hybrid calculations, one could use LDA or GGA functionals to test the supercell size convergence.
Note however, that this is a two-stage approach: in the first instance one must test that within the same supercell size the two methods yield similar trends. %there is qualitative agreement between the two methods.
%This is a two stage approach: in the first instance one perform the simulations in both levels of theory with the same simulation size. If qualitative agreement is found, then the lower level of theory may be used to 
%Thus, if a reliable force-field potential exist for the system of interest, one could perform force-field simulations using the same supercell as the DFT calculations; if there is qualitative agreement for that size, then convergence with respect to supercell size may be carried out with the force-field potentials. Similarly, for hybrid calculations, if there is qualitative agreement between the hybrid functional and LDA or GGA (when simulated int he same supercell), these functionals may be used to test supercell convergence.
%Thus one may use a lower level of theory method to perform the convergence test, provided there is qualitative agreement between the two methods when comparing results from the same (small) simulation cell.
Quantitative agreement is not expected, but any qualitative agreement between the two methods observed in the small simulation size is expected to be preserved in larger simulations cells (as shown in the current work).
Note that this method cannot be extended to electronic self-interaction, as these are highly sensitive to the description of the electronic states~\cite{Centoni2005}.
\item The alternative approach is to use a range-dependent constraint method with $R_\text{relax}=\frac{L_\text{min}}{2}$, as illustrated in Figure~\ref{fig:seldyn}, to test whether there is qualitative agreement between the fully relaxed simulations and those where the strain fields were not allowed to transfer across the PBC. Incidentally, this method also significantly reduces the computational cost of energy minimisation on account of the reduced internal degrees of freedom. The size of the discrepancy may be considered as an indicator of the extent to which strain extends beyond the PBCs.
\end{enumerate}

\section{Conclusions}
Electrostatic self-interactions have long been thought to be the dominant source of finite size effects in ionic materials. Here we have shown that even in ionic compounds, elastic self-interactions are non-negligible, causing qualitative and quantitative changes to the energy and structure of defects. This finite size effect is the source of numerous inaccurate reports in the \emph{ab-initio} literature regarding the defect stability of many important functional oxides including CeO$_2$, ThO$_2$ and UO$_2$.
The magnitude of the elastic self-interaciton is such that it warrants the need for simulations sizes larger than the widely used 96-atoms supercell for cubic oxide compounds. In addition, we show that the spurious self-interaction cannot be countered by simple linear elastic theory approaches. 

We considered the exemplar cases of charge-neutral Schottky clusters in fluorite-structured oxides and perform DFT and force-field simulations in supercells of increasing size up to 1500 atoms. 
Supercells of 96 atoms were not sufficiently large to capture the most stable neutral Schottky clusters compared to larger cells. 
%We observe a change in the most stable neutral cluster configuration for sufficiently large supercells.
This behaviour is observed also for non-oxide ionic compounds with related structures (prototypical fluorite CaF$_2$, and antifluorite Be$_2$C), but was not observed in iso-structural metallic compounds. This phenomenon is also insensitive to the level of theory used (EAM, DFT, DFT+U).

We provide evidence that the finite size effect is not due to electrostatic self-interaction (accounting for charge-charge, charge-dipole and dipole-dipole interactions), nor electronic effects (owing to the presence of the self-interaction in force-field calculations with fixed charges) instead it is caused by spurious interaction of strain fields across PBC. This is confirmed through the use of range-constrains, where selected atoms were fixed to the perfect lattice sites, thereby artificially truncating the strain field before the PBC. In these simulations  the correct order of defect energies was restored.
Surprisingly, the spurious self-interaction energy cannot be accounted for, and therefor corrected with, linear elastic theory. The failure of linear elastic theory is attributed, through careful analysis of the displacement fields, to the complex relaxation patterns observed in the ionic compounds, akin to optical modes. 

Our findings also resolve a long-standing discrepancy between DFT and force-field simulations regarding the ground state of neutral defect clusters in actinide oxides. This had typically been attributed to the simplified atomic interactions of the empirical force-field methods, but we show that good agreement is in fact obtained when comparing across the same supercell sizes. The fact that the phenomenon is observed at all levels of theory can be exploited to the practitioners' advantage, using computationally simpler methods to test the degree of elastic self-interaction in the system of interest.

Although we have focussed primarily on bound Schottky defects, the findings are likely relevant to other neutral clusters and their migration pathways, such as $\left\{2\text{Y}_\text{Zr}^{\prime}:V_\text{O}^{\boldsymbol{\cdot\cdot}}\right\}^{\times}$ and $\left\{2\text{Gd}_\text{Ce}^{\prime}:V_\text{O}^{\boldsymbol{\cdot\cdot}}\right\}^{\times}$ in yttria-stabilised-zirconia and gadolinia-doped ceria. %Given the elastic origin of the finite size effect, it is expected to have implications for non-ionic compounds, as well as compounds with non-cubic structure (e.g.\ tetragonal and monoclinic ZrO$_2$ and HfO$_2$).
Rather than just being considered a spurious size effect for dilute limit calculations, the elastic effects discussed here also apply to real materials with high concentrations of Schottky defects, providing an insight into the interaction between two or more Schottky clusters that come within 10--16\si{\angstrom} of each other. This may be an important factor in the nucleation of voids in nuclear fuels and SOCF. 

%We find that supercells with minimum dimensions of $16\times16\times16$ \si{\angstrom^3} are required for quantitatively accurate results of Schottky clusters in fluorite-structure compounds.
%We have performed DFT simulations with up to 1500 atoms, but this is not always computationally practicable, especially with more advanced \emph{ab-initio} methods. Thus, we proposed a two-region scheme that retains the qualitative description of the defect even when simulated in insufficiently small supercells.
%To address the computational difficulties associated with the large system sizes used here, we proposed a two-region scheme that retains the qualitative description of the defect even when simulated in small supercells. In addition, the scheme reduces the computational time of energy minimisation, thus potentially enabling the use of larger supercells, thereby yielding qualitatively more accurate results.

The current work should serve as a cautionary tale for future simulations of all ionic systems, especially those where computational requirements dictate the use of small supercell sizes. Whilst proving that 96-atom supercells are inadequate for point defect analysis in fluorite-structured oxides, every solid state system would have different supercell requirements. In the current paper we provide two computationally efficient methods to ascertain whether elastic finite-size effects are significant for the system of interest: 1) comparison with lower level of theory (as exemplified here using force-field CRG potentials), 2) comparison with range-constrained simulations (where selected atoms have been fixed to the perfect lattice sites to hamper propagation of strain fields).

\begin{acknowledgments}
The authors would like to acknowledge S.~T.~Murphy, N.~Kuganathan, M.~J.~D.~Rushton and R.~W.~Grimes for the useful discussion. M.~W.~D.~Cooper is funded by the US Department of Energy, Office of Nuclear Energy, Nuclear Energy Advanced Modeling Simulation (NEAMS) program. Los Alamos National Laboratory, an affirmative action/equal opportunity employer, is operated by Los Alamos National Security, LLC, for the National Nuclear Security Administration of the U.S.~Department of Energy under Contract No.~DE-AC52- 06NA25396. 
P.~A.~Burr acknowledges the Australian Nuclear Science and Technology Organisation and the Tyree Foundation for their charitable financial support.
\end{acknowledgments}

%Given the elastic origin of these size effects, rather than electrostatic, we expect this work to also have implications beyond ionic systems to the modelling of metallic and covalent compounds. 
%However, since no displacement is observed, it is not possible to measure and counter this positive energy contribution through a dipole or elastic greens function method.
%A quantitative description of the atomic displacements (again for CeO$_2$) is presented in Figure~\ref{fig:displacements}, for each supercell size and Schottky configuration and. 

%%%%%%%%%%%%%%%%%       BIBLIOGRAPHY        %%%%%%%%%%%%%%%%%%%%%

%\section*{References}

\bibliographystyle{apsrev4-1}

%\begin{thebibliography}
\bibliography{/Users/pab07/Documents/papers/MyCollection}
%\bibliography{/Users/pab07/Documents/papers/library}			%auto-sync
%\end{thebibliography}

\end{document}